# Distributed Equilibrium-Learning for Power Network Voltage Control With a Locally Connected Communication Network

Kaiqing Zhang, Wei Shi, Hao Zhu, and Tamer Başar

*Abstract*— In current power distribution systems, one of the most challenging operation tasks is to coordinate the network-wide distributed energy resources (DERs) to maintain the stability of voltage magnitude of the system. This voltage control task has been investigated actively under either distributed optimization-based or local feedback control-based characterizations. The former architecture requires a *strongly-connected* communication network among all DERs for implementing the optimization algorithms, a scenario not yet realistic in most of the existing distribution systems with under-deployed communication infrastructure. The latter one, on the other hand, has been proven to suffer from loss of network-wide operational optimality. In this paper, we propose a game-theoretic characterization for *semi-local* voltage control with only a locally connected communication network. We analyze the existence and uniqueness of the generalized Nash equilibrium (GNE) for this characterization and develop a fully distributed equilibrium-learning algorithm that relies on only *neighbor-to-neighbor* information exchange. Provable convergence results are provided along with numerical tests which corroborate the robust convergence property of the proposed algorithm.

## I. INTRODUCTION

Increasing penetration of distributed energy resources (DERs) has significantly facilitated the regulation of voltage level in power distribution systems. By varying the reactive power (VAR) injected into the system, the DERs can be coordinated to attain the desirable voltage profile in a timely fashion. This problem can be formulated as a network-wide optimal power flow (OPF) one. For recent efforts based on this formulation, we refer to [1], [2], [3], [4], and the references therein. Central/distributed optimization algorithms such as primal-dual and ADMM have been advocated to solve the OPF problem. A majority of these algorithms require a *strongly-connected* communication network at high quality in order to diffuse the local information (either decision variables or measurement) across all DERs in the network. Therefore, the limited communication capabilities available to distribution systems challenge the practical implementation of these algorithms.

The design of resource coordination algorithms for DERs has to address the status quo of the communication and networking technologies in current distribution systems. Several competing solutions, such as fiber optic cable, point-to-point

This work is partially supported by the National Science Foundation under Award Number ECCS-1610732 and the Power Systems Engineering Research Center (PSERC) Project S-70. K. Zhang and T. Başar are with the Coordinated Science Laboratory, University of Illinois at Urbana-Champaign ({kzhang66, basar1}@illinois.edu). W. Shi is with Arizona State University (wilbur.shi@asu.edu). H. Zhu is with Department of Electrical and Computer Engineering at the University of Texas at Austin (haozhu@utexas.edu).

microwave, and powerline carrier, are all subject to their own limitations such as bandwidth constraints, unacceptable delay, and high deployment cost [5], [6]. Some recent work has accounted for these communication limitations in designing optimization-based voltage control algorithms. For example, [2] considers a hybrid voltage control strategy that is cognizant to the instantaneous unavailability of communication links. In [7], distributed voltage control algorithms using only quantized communication between neighboring buses are developed to adapt to the bandwidth constraints.

On the other hand, several *local* control schemes have been developed to allow for the need of no information exchange in real-time. As advocated in [8], [9], [10], DERs can perform *feedback* control using only local voltage measurement at no communications, even under asynchronous updates [11]. Nonetheless, the local schemes have been proven to make *myopic decisions* and lead to loss of optimality in voltage control performance [8], [9]. It has been argued in [12] that the trade-off exists between performance optimality and communication complexity for the voltage control problem, as in a variety of control applications.

To better characterize the performance-communication trade-off, this paper proposes a *semi-local* control scheme under a *locally connected communication network*. In particular, we consider the scenario where DERs are partitioned into several *communication areas*, where information can be exchanged within each area. This scenario nicely generalizes the scenario for local control schemes where each DER itself is a communication area, as well as the scenario under a strongly-connected communication network with a single area. Moreover, it fills the gaps in between the two special scenarios, as the achievable performance under locally connected communication network characterizes the value of communication links. Last but not least, this scenario is extremely useful for the design of networked micro-grids [13], where physically connected micro-grids are owned and controlled by different entities or operators. Due to competition or privacy concerns, there is very limited information exchange among the micro-grids.

To pursue the semi-local control scheme, we develop a game-theoretic characterization for this problem. The areas controlled by different operators are modeled as players in a strategic game and only considers self interest due to lack of communication with each other. The contribution of the present work is three-fold: i) we analyze the existence and provide uniqueness conditions for the generalized Nash equilibrium (GNE) under the game-theoretic characterization; ii) we propose a fully distributed equilibrium-learning algorithm for voltage control that requires only neighbor-to-neighbor communication between DERs in the same area; iii) we prove the convergence of the algorithm under certain monotonicity conditions and also corroborate its superior convergence property numerically.

It is worth mentioning that the proposed equilibrium-learning algorithm should be of independent interest for solving the GNE problem with similar communication constraints. It is designed based on the inexact alternating direction method of multipliers (inexact-ADMM) for solving distributed optimization problems [14]. As pointed out by the survey paper [15], very few algorithms have been proposed for learning the GNE based on Lagrangian relaxation of the coupling constraints as we propose here, letting alone the ADMM-typed methods. One recent work [16] develops an augmented Lagrangian-typed method for finding GNEs; however, its implementation is not fully distributed within the area as in our setting. Thanks to the availability of real-time measurements in power networks, our algorithm does not rely on information from other areas (players) of the game. Although convergence at $O(1/t)$ rate is proved only under certain conditions, the proposed learning algorithm exhibits superior convergence property even when these conditions are violated, as demonstrated by numerical simulations.

## II. Modeling and Problem Formulation

In this section, we introduce the modeling of power flow in distribution networks and the voltage control problem with only a locally connected communication network.

### A. Voltage control problem

Consider a single-phase power distribution network with a tree-topology graph denoted by $(\mathcal{N}, \mathcal{E})$, where $\mathcal{N} := \{0, \cdots, N\}$ is the set of buses with bus-0 as the feeder and $\mathcal{E} := \{(i,j), \forall i, j \in \mathcal{N}\}$ is the set of line segments. We denote bus-0 as the reference bus and all the other buses as a set $\mathcal{N}_p := \mathcal{N}/\{0\}$ that has controllable DERs providing reactive power injection. The voltage magnitude $v_j$ and the reactive power injection $q_j$ at each bus-$j$, $\forall j \in \mathcal{N}_p$, are concatenated in $\mathbf{v} \in \mathbb{R}^N$ and $\mathbf{q} \in \mathbb{R}^N$, respectively. Note that $v_0$ is the reference bus voltage for the point of common coupling (PCC). Using the so-termed *LinDisFlow model* [17], a linearized branch flow model for distribution networks, we establish the relation between the VAR injection $\mathbf{q}$ and the voltage profile $\mathbf{v}$ as in [2]

$$\mathbf{Bv} = \mathbf{q} + \mathbf{w} \tag{1}$$

where $\mathbf{w}$ is a constant vector that captures the system operating point, $\mathbf{B}$ is the reduced[1] weighted Laplacian matrix of $(\mathcal{N}, \mathcal{E})$, satisfying $B_{ij} = B_{ji} = 0, \forall (i,j) \notin \mathcal{E}$. This sparsity structure of $\mathbf{B}$ will facilitate the design of distributed algorithms for voltage regulation using only neighbor-to-neighbor communications.

To regulate the system voltage profile to a desired one, i.e., $\mathbf{v} \to \boldsymbol{\mu}$, while considering the cost and limits of

[1]By saying *reduced*, we mean that the row and the column corresponding to bus-0 are removed in the Laplacian matrix.

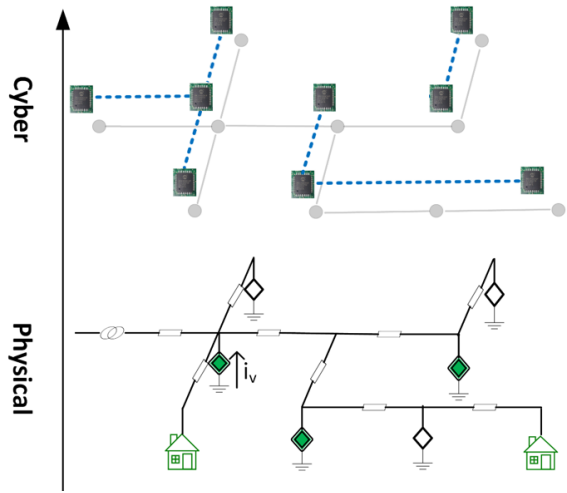

Fig. 1. A schematic diagram for current distribution system with a locally connected communication network illustrated as a cyber-physical system.

VAR provision, we formulate the following voltage control problem denoted as $\mathcal{P}_0$

$$\min_{\mathbf{v},\mathbf{q}} \quad \frac{\gamma}{2}\|\mathbf{v}-\boldsymbol{\mu}\|_2^2 + \sum_{j=1}^N C_j(q_j) \tag{2a}$$

$$s.t. \quad \mathbf{Bv} = \mathbf{q} + \mathbf{w} \tag{2b}$$

$$q_j \in \mathcal{Q}_j, \forall j \in \mathcal{N}_p \tag{2c}$$

where $\mathcal{Q}_j := [\underline{q}_j, \bar{q}_j]$ denotes the box-constrained feasible set for VAR provision, $\gamma \geq 0$ is a parameter that balances the voltage regulation and the VAR provision cost $C_j(q_j)$ in the objective. We make a standard assumption on $C_j(q_j)$.

**Assumption 1.** *The cost functions $C_j(q_j)$ are convex and continuously-differentiable over $q_j \in \mathcal{Q}_j, \forall j \in \mathcal{N}_p$.*

### B. Locally connected communication network

As shown in Fig. 1, we consider a communication network $(\mathcal{N}, \mathcal{E}_c)$ constructed on top of the distribution network $(\mathcal{N}, \mathcal{E})$ with the same set of buses[2]. This two-layer network is usually viewed as a cyber-physical system. In current distribution systems, however, the communication links are relatively scarce so that the physically connected buses are not necessarily all connected over the communication network. Our focus is to analyze the performance limits and design admissible voltage control algorithms when there is such restricted communication.

Suppose there are $K$ connected *communication areas* in $(\mathcal{N}, \mathcal{E}_c)$ and let $\mathcal{K} = \{1, \cdots, K\}$. Let $\mathcal{K}_k$ denote the set of buses within the $k$-th communication area. Each bus can merely exchange measurement or decisions with neighboring buses in the same area. Hence for each area-$k$, the relation between the VAR injection $\mathbf{q}_k \in \mathbb{R}^{|\mathcal{K}_k|}$ and the voltage profile $\mathbf{v}_k \in \mathbb{R}^{|\mathcal{K}_k|}$ is partitioned from (1) as,

$$\mathbf{B}_{k,k}\mathbf{v}_k = \mathbf{q}_k + \mathbf{w}_k - \mathbf{B}_{k,-k}\mathbf{v}_{-k} \tag{3}$$

[2]Mathematically speaking, this leads to the fact that $(\mathcal{N}, \mathcal{E}_c)$ is a subgraph of $(\mathcal{N}, \mathcal{E})$.

where the subscript $-k$ represents the indices of buses that are not in communication area-$k$, i.e., $\{j|j \notin \mathcal{K}_k\}$, $\mathbf{B}_{k,k}$ and $\mathbf{B}_{k,-k}$ are thus submatrices of $\mathbf{B}$ with proper dimensions. Note that the submatrix $\mathbf{B}_{k,k}$ inherits the sparsity structure as the matrix $\mathbf{B}$. Denoting $\tilde{\mathbf{B}} := \mathrm{diag}\{\mathbf{B}_{k,k}\}$ as a block diagonal matrix with $\mathbf{B}_{k,k}$ on its diagonal, we immediately have the following lemma.

**Lemma 1.** *Both matrices $\mathbf{B}$ and $\tilde{\mathbf{B}}$ are positive definite (PD) and thus invertible.*

**Proof.** The positive definiteness of $\mathbf{B}$ follows directly from [9, Prop. 1]. Hence all the block submatrices $\mathbf{B}_{k,k}, \forall k \in \mathcal{K}$ are PD, and so does the block diagonal matrix $\tilde{\mathbf{B}}$. ∎

### III. GAME CHARACTERIZATION WITH LOCAL COMMUNICATIONS

In this section, we consider a game theoretic characterization of the voltage control problem with only a locally connected communication network.

With non-strongly connected communications, the global problem $\mathcal{P}_0$ can hardly be attacked directly using distributed optimization algorithms as [18], [19] since information cannot spread over all buses. Alternatively, the buses of each communication area can only take care of the voltage mismatch and VAR injection within that area, though they are physically coupled with buses of other areas. This characteristic makes the problem fall under the realm of strategic games. Consider a game $\mathcal{G} = \langle \mathcal{K}, \{\mathcal{A}_k\}_{k \in \mathcal{K}}, \{U_k\}_{k \in \mathcal{K}} \rangle$ with $K$ players whose *feasible sets* $\mathcal{A}_k, \forall k \in \mathcal{K}$ are defined as follows

$$\mathcal{A}_k := \{(\mathbf{v}_k, \mathbf{q}_k) | \underline{\mathbf{q}}_k \leq \mathbf{q}_k \leq \bar{\mathbf{q}}_k, \qquad (4)$$
$$\mathbf{B}_{k,k}\mathbf{v}_k = \mathbf{q}_k + \mathbf{w}_k - \mathbf{B}_{k,-k}\mathbf{v}_{-k}\}.$$

Let $\mathbf{a}_k := (\mathbf{v}_k, \mathbf{q}_k)$ denote the decision variable of area-$k$. Note that $\mathcal{A}_k$ is determined by the action of other areas, denoted as $\mathbf{a}_{-k}$, i.e., $\mathcal{A}_k = \mathcal{A}_k(\mathbf{a}_{-k})$. Let $U_k : \mathcal{A}_k \to \mathbb{R}$ denote the *payoff function* for area-$k$. Similarly to (2a), each area aims to minimize its own operational cost with the form[3]:

$$U_k(\mathbf{a}_k) = \frac{\gamma}{2}\|\mathbf{v}_k - \boldsymbol{\mu}_k\|_2^2 + \sum_{j \in \mathcal{K}_k} C_j(q_j). \qquad (5)$$

Hence, the voltage control problem for each area-$k$ can be formulated as

$$\min_{\mathbf{a}_k} U_k(\mathbf{a}_k) \quad s.t. \quad \mathbf{a}_k \in \mathcal{A}_k(\mathbf{a}_{-k}), \quad \forall k \in \mathcal{K} \qquad (6)$$

The problem (6) is commonly categorized as a generalized Nash equilibrium problem (GNEP), in which each player has its own objective and action while its feasible set is dependent on the actions of other players. The solution to the GNEP, i.e., the generalized Nash equilibrium, characterizes the performance limit of voltage control for such a locally connected communication network. In general, however, it is not straightforward to establish the existence and uniqueness of the GNE [15]. Here due to the special structure of $\mathcal{G}$, we can establish the following results on the existence and uniqueness of the GNE based on the theory of variational inequality.

---

[3]Note here we use the convention that players *minimize* (not maximize) their cost (payoff) functions as in [20].

**Theorem 1.** *[Existence of the GNE] The set of the GNE for $\mathcal{G}$ is nonempty and compact.*

**Proof.** Under Assumption 1, the GNEP for each area-$k$ is a convex optimization problem for given other areas' decision $\mathbf{a}_{-k}$. Hence the Karush-Kuhn-Tucker (KKT) conditions equivalently characterize the solution to each GNEP since the non-empty interior points in $\mathcal{A}_k(\mathbf{a}_{-k})$ ensure the Slater's condition to hold [21] for each subproblem (6). We thus obtain the equilibrium conditions (EC) of the game $\mathcal{G}$, i.e., $\forall k \in \mathcal{K}$,

$$\begin{cases} \gamma(\mathbf{v}_k - \boldsymbol{\mu}_k) + \mathbf{B}_{k,k}^T \boldsymbol{\theta}_k = \mathbf{0} & (7a) \\ \nabla \mathbf{C}_k(\mathbf{q}_k) - \boldsymbol{\theta}_k + \bar{\boldsymbol{\eta}}_k - \underline{\boldsymbol{\eta}}_k = \mathbf{0} & (7b) \\ \bar{\boldsymbol{\eta}}_k^T \bar{\mathbf{q}}_k = 0, \underline{\boldsymbol{\eta}}_k^T \underline{\mathbf{q}}_k = 0, \bar{\boldsymbol{\eta}}_k \geq \mathbf{0}, \underline{\boldsymbol{\eta}}_k \geq \mathbf{0} & (7c) \\ \underline{\mathbf{q}}_k \leq \mathbf{q}_k \leq \bar{\mathbf{q}}_k, \mathbf{B}_{k,k}\mathbf{v}_k = \mathbf{q}_k + \mathbf{w}_k - \mathbf{B}_{k,-k}\mathbf{v}_{-k} & (7d) \end{cases}$$

where $\nabla \mathbf{C}_k(\mathbf{q}_k) := [\nabla C_j(q_j)]_{j \in \mathcal{K}_k}$ and $\bar{\boldsymbol{\eta}}_k, \underline{\boldsymbol{\eta}}_k$ and $\boldsymbol{\theta}_k$ are the multipliers. Equations (7a) and (7b) are the stationarity conditions with respect to (w.r.t.) $\mathbf{v}_k$ and $\mathbf{q}_k$, respectively. Equation (7c) represents the complementary slackness and dual feasibility conditions, and equation (7d) represents the primal feasibility conditions. By eliminating the variable $\mathbf{v}_k$ using $\mathbf{q}_k$, we obtain the EC in the following compact form

$$\begin{cases} \gamma \tilde{\mathbf{B}}^{-1}(\mathbf{B}^{-1}\mathbf{q} + \bar{\boldsymbol{\mu}}) + \nabla \mathbf{C}(\mathbf{q}) + \bar{\boldsymbol{\eta}} - \underline{\boldsymbol{\eta}} = \mathbf{0} & (8a) \\ \bar{\boldsymbol{\eta}}^T(\mathbf{q} - \bar{\mathbf{q}}) = 0, \underline{\boldsymbol{\eta}}^T(\mathbf{q} - \underline{\mathbf{q}}) = 0, \bar{\boldsymbol{\eta}} \geq \mathbf{0}, \underline{\boldsymbol{\eta}} \geq \mathbf{0} & (8b) \\ \underline{\mathbf{q}} \leq \mathbf{q} \leq \bar{\mathbf{q}} & (8c) \end{cases}$$

where $\bar{\boldsymbol{\mu}} := \mathbf{B}^{-1}\mathbf{w} - \boldsymbol{\mu}$ is constant, and $\nabla \mathbf{C}(\mathbf{q})$ is a concatenation of $\nabla \mathbf{C}_k(\mathbf{q}_k)$. Interestingly, the EC (8) coincides with the KKT system of a variational inequality $\mathrm{VI}(\mathcal{Q}, \boldsymbol{\Phi})$, where $\mathcal{Q} := \prod_{j \in \mathcal{N}_p} \mathcal{Q}_j$ is a convex set and $\boldsymbol{\Phi} : \mathcal{Q} \to \mathbb{R}^N$ is a mapping defined as

$$\boldsymbol{\Phi}(\mathbf{q}) := \gamma \tilde{\mathbf{B}}^{-1}(\mathbf{B}^{-1}\mathbf{q} + \bar{\boldsymbol{\mu}}) + \nabla \mathbf{C}(\mathbf{q}). \qquad (9)$$

The solution set to the $\mathrm{VI}(\mathcal{Q}, \boldsymbol{\Phi})$ is equivalently the solution of the KKT system (8) (see [22, Prop. 1.3.4]), which is nonempty and compact since $\mathcal{Q}$ is convex and compact and $\boldsymbol{\Phi}$ is continuous by Assumption 1 (see [22, Corollary 2.2.5]). This completes the proof. ∎

One advantage of this connection to the $\mathrm{VI}(\mathcal{Q}, \boldsymbol{\Phi})$ is to elicit the conditions for the uniqueness of the GNE, for more discussions see [22]. One of the most commonly used conditions for global uniqueness of the VI's solution is as follows [20].

**Lemma 2.** *The GNE for $\mathcal{G}$ is unique if the mapping $\boldsymbol{\Phi} : \mathcal{Q} \to \mathbb{R}^N$ defined in (9) is strongly monotone, i.e., $\forall \mathbf{q}^1, \mathbf{q}^2 \in \mathcal{Q}$, there exists a constant $m > 0$ such that*

$$(\mathbf{q}^1 - \mathbf{q}^2)^T[\boldsymbol{\Phi}(\mathbf{q}^1) - \boldsymbol{\Phi}(\mathbf{q}^2)] \geq m\|\mathbf{q}^1 - \mathbf{q}^2\|_2^2.$$

Another interesting advantage of the connection to the VI

theory is that some algorithms for solving the VI can be used to seek the GNE. For example, the basic *projection algorithm* [23, Chapter 3.5.2], which performs the projected gradient-play $\mathbf{q}^{(t+1)} = \mathbb{P}_\mathcal{Q}[\mathbf{q}^{(t)} - \epsilon \mathbf{\Phi}(\mathbf{q}^{(t)})]$[4] on the feasible set $\mathcal{Q}$, has its fixed point as the GNE provided that the mapping $\mathbf{\Phi}$ is strongly monotone and the step-size $\epsilon$ is small enough. However, from (9), the evaluation of $\mathbf{\Phi}$ requires a fully connected communication within each area since the block matrices $\mathbf{B}_{k,k}^{-1}$ on the diagonal of $\tilde{\mathbf{B}}^{-1}$ are usually dense (due to the sparsity structure of its inverse, $\mathbf{B}_{k,k}$). Hence this equilibrium-learning approach may not be implementable under our setting with local communications as illustrated in Fig. 1. This motivates us to develop a fully distributed equilibrium-learning algorithm that requires only neighbor-to-neighbor communications in the next section.

## IV. Distributed Equilibrium-learning for Voltage Control

In this section, we propose a fully distributed voltage control algorithm to achieve the GNE while requiring only neighbor-to-neighbor information exchange within each communication area.

### A. The GNE and VI in primal-dual space

As discussed in Section III, the solution to the VI$(\mathcal{Q}, \mathbf{\Phi})$ aligns with the equilibrium of the voltage control under local communications. However, due to the general dense structure of $\tilde{\mathbf{B}}^{-1}$, most algorithms for solving VI$(\mathcal{Q}, \mathbf{\Phi})$, as summarized in [22, Chapters 10 and 12], can not be implemented in a fully distributed fashion. To exploit the sparsity structure in $\tilde{\mathbf{B}}$, dualization of the equality constraint in (3) seems to be necessary. A Lagrangian dual based algorithm has been proposed and studied in reference [2] under the condition of having a globally connected communication network. This algorithm updates iterates in both primal and dual spaces $(\mathbf{v}, \mathbf{q}, \boldsymbol{\theta}) \in \Omega := \mathbb{R}^N \times \mathcal{Q} \times \mathbb{R}^N$ (unlike the primal-based algorithms such as gradient-projection which only perform updates in the primal space $\mathbf{q} \in \mathcal{Q}$). Here, the variable $\boldsymbol{\theta}$ is the dual variable associated with the constraint (3).

Based on this observation, we first establish the connection between the GNE and a *VI in the primal-dual variable space*. For ease of notation, we denote that $\forall k \in \mathcal{K}$,

$$f_k(\mathbf{v}_k) := \frac{\gamma}{2} \|\mathbf{v}_k - \boldsymbol{\mu}_k\|_2^2 \text{ and } h_k(\mathbf{q}_k) := \sum_{j \in \mathcal{K}_k} C_j(q_j),$$

$$f(\mathbf{v}) := \sum_{k \in \mathcal{K}} f_k(\mathbf{v}_k) \text{ and } h(\mathbf{q}) := \sum_{k \in \mathcal{K}} h_k(\mathbf{q}_k).$$

Hence given any $\mathbf{a}_{-k}$, to solve each GNEP (6) is to find a point $\boldsymbol{\omega}_k^* := (\mathbf{v}_k^*, \mathbf{q}_k^*, \boldsymbol{\theta}_k^*)^T$ such that $\forall \mathbf{v}_k \in \mathbb{R}^{|\mathcal{K}_k|}, \forall \mathbf{q}_k \in \mathcal{Q}_k$,

$$\begin{cases} (\mathbf{v}_k - \mathbf{v}_k^*)^T (\nabla f_k(\mathbf{v}_k^*) + \mathbf{B}_{k,k}^T \boldsymbol{\theta}_k^*) \geq 0, & \text{(10a)} \\ (\mathbf{q}_k - \mathbf{q}_k^*)^T (\nabla h_k(\mathbf{q}_k^*) - \boldsymbol{\theta}_k^*) \geq 0, & \text{(10b)} \\ \mathbf{B}_{k,k} \mathbf{v}_k^* = \mathbf{q}_k^* + \mathbf{w}_k - \mathbf{B}_{k,-k} \mathbf{v}_{-k}. & \text{(10c)} \end{cases}$$

---

[4]$\mathbb{P}_\mathcal{Q}[\cdot]$ denotes the projection operator onto the set $\mathcal{Q}$.

By concatenating (10) for all $k \in \mathcal{K}$, the GNEP is equivalent to finding an $\boldsymbol{\omega}^* := (\mathbf{v}^*, \mathbf{q}^*, \boldsymbol{\theta}^*)^T \in \Omega$ such that

$$\text{VI}(\Omega, \mathbf{F}) : (\boldsymbol{\omega} - \boldsymbol{\omega}^*)^T \mathbf{F}(\boldsymbol{\omega}^*) \geq 0, \forall \boldsymbol{\omega} \in \Omega, \quad (11)$$

where

$$\boldsymbol{\omega} := \begin{bmatrix} \mathbf{v} \\ \mathbf{q} \\ \boldsymbol{\theta} \end{bmatrix}, \mathbf{F}(\boldsymbol{\omega}) := \begin{bmatrix} \nabla f(\mathbf{v}) + \tilde{\mathbf{B}}^T \boldsymbol{\theta} \\ \nabla h(\mathbf{q}) - \boldsymbol{\theta} \\ -\rho \tilde{\mathbf{B}}(\mathbf{v} - \mathbf{B}^{-1}\mathbf{q} - \mathbf{B}^{-1}\mathbf{w}) \end{bmatrix}.$$

To facilitate our proof later, the power flow constraint (10c) is equivalently rewritten as $\rho \tilde{\mathbf{B}}(\mathbf{v} - \mathbf{B}^{-1}\mathbf{q} - \mathbf{B}^{-1}\mathbf{w}) = \mathbf{0}$ in the definition of $\mathbf{F}$ with any fixed $\rho > 0$. This is due to Lemma 1 that both $\tilde{\mathbf{B}}$ and $\mathbf{B}$ are invertible.

**Remark 1.** Several algorithms have been proposed in [22] to solve general VI problems. To guarantee the global convergence of the algorithm, the mapping $\mathbf{F}$ is assumed to have some form of monotonicity, see Chapter 12 in [22]. Among these assumptions on monotonicity, *pseudo-monotone* is the weakest one, under which the *extra-gradient* (EG) algorithm [22], to the best of our knowledge, is the only one that has theoretical guarantee of global convergence to the VI solution. Moreover, since the sparse matrix $\tilde{\mathbf{B}}$ is used to evaluate $\mathbf{F}$, the gradient of the VI, the EG algorithm admits a distributed implementation. Therefore, we will use it as the benchmark algorithm in our simulations. In fact, the poor convergence speed of the EG algorithm shown later motivates us to propose the following novel primal-dual algorithm for equilibrium-learning, although its convergence result can so far only be established on certain monotonicity assumptions as we will show shortly.

### B. Algorithm design

The equilibrium-learning algorithm is inspired by primal-dual algorithms for solving convex optimization problems [24], especially the ADMM algorithm [14]. We also incorporate the *feedback control scheme* that exploits real-time measurement of voltage magnitude as in [2], to enable a distributed implementation of the algorithm.

First for each iteration $t$, define

$$\mathbf{g}^{(t)} := \mathbf{B}_{k,k}^T \big(\mathbf{B}_{k,k} \mathbf{v}_k^{(t)} - \mathbf{q}_k^{(t)} - \check{\mathbf{w}}_k^{(t)} + \frac{1}{\rho} \boldsymbol{\theta}_k^{(t)}\big), \quad (12)$$

where $\rho > 0$ is a parameter of the algorithm to be set. $\check{\mathbf{w}}_k^{(t)}$ is a measurable quantity defined as

$$\check{\mathbf{w}}_k^{(t)} := \mathbf{B}_{k,k} \check{\mathbf{v}}_k^{(t)} - \mathbf{q}_k^{(t)},$$

where $\check{\mathbf{v}}^{(t)}$ is the measurement of voltage magnitude when the VAR $\mathbf{q}^{(t)}$ is injected, determined by the power flow equation (1). This way, the decision related to other areas at time $t$, i.e., the term $\mathbf{w}_k - \mathbf{B}_{k,-k} \mathbf{v}_k^{(t)}$ in (3), can be obtained through this *feedback* measurement without any communication across areas.

Now we are ready to present the update steps of the proposed algorithm, i.e., $\forall k \in \mathcal{K}$,

$$\mathbf{v}_k^{(t+1)} = \arg\min_{\mathbf{v}_k} \frac{\gamma}{2}\|\mathbf{v}_k - \boldsymbol{\mu}_k\|_2^2$$
$$+ \rho \left\langle \mathbf{g}^{(t)}, \mathbf{v}_k - \mathbf{v}_k^{(t)} \right\rangle + \frac{\beta}{2}\|\mathbf{v}_k - \mathbf{v}_k^{(t)}\|_2^2 \quad (13a)$$

$$\mathbf{q}_k^{(t+1)} = \arg\min_{\mathbf{q}_k \in \mathcal{Q}_k} \sum_{j \in \mathcal{K}_k} C_j(q_j)$$
$$+ \frac{\rho}{2}\|\mathbf{B}_{k,k}\mathbf{v}_k^{(t+1)} - \mathbf{q}_k - \check{\mathbf{w}}_k^{(t)} + \frac{1}{\rho}\boldsymbol{\theta}_k^{(t)}\|_2^2 \quad (13b)$$

$$\boldsymbol{\theta}_k^{(t+1)} = \boldsymbol{\theta}_k^{(t)} + \rho(\mathbf{B}_{k,k}\mathbf{v}_k^{(t+1)} - \mathbf{q}_k^{(t+1)} - \check{\mathbf{w}}_k^{(t+1)}) \quad (13c)$$

where $\beta > 0$ is also a parameter to be tuned. The updates (13) originate from the ADMM update for solving the sub-problem GNEP (6) of each area-$k$. The first step (13a) exploits a linearized/inexact minimization so that this step can be implemented using only neighbor-to-neighbor communications. In particular, the updates at each bus can be written in closed form, i.e., $\forall j \in \mathcal{K}_k$ and $\forall k \in \mathcal{K}$,

$$v_j^{(t+1)} = \frac{\gamma \mu_j - \sum_{i \in \mathcal{N}_c^j} B_{ji}(2\theta_i^{(t)} - \theta_i^{(t-1)}) + \beta v_j^{(t)}}{\gamma(1+\beta)} \quad (14a)$$

$$q_j^{(t+1)} = \mathrm{Sol}_j \left[ \nabla C_j(q_j) + \rho(q_j - q_j^{(t)} - \frac{1}{\rho}\theta_j^{(t)}) \right.$$
$$\left. - \rho \sum_{i \in \mathcal{N}_c^j} B_{ji}(v_j^{(t+1)} - \check{v}_j^{(t)}) \right] \quad (14b)$$

$$\theta_j^{(t+1)} = \theta_j^{(t)} + \rho \sum_{i \in \mathcal{N}_c^j} B_{ji}(v_j^{(t+1)} - \check{v}_j^{(t+1)}) \quad (14c)$$

where $\mathcal{N}_c^j$ denotes the neighboring buses of bus-$j$ over the communication network $(\mathcal{E}, \mathcal{N}_c)$, the operator $\mathrm{Sol}_j[g(q_j)]$ over a continuous function $g(q_j)$ finds the solution of $g(q_j) = 0$ and then projects it onto the feasible set $\mathcal{Q}_j$. Note that by Assumption 1, the optimization in step (13b) is strongly convex w.r.t. $q_j$, which ensures the uniqueness of the zero of the mapping in the square brackets of (14b). All three updates in (14) require only local information exchange with the neighboring buses, respecting the locally connected communication topology considered here.

*C. Convergence analysis*

Here we will show that the updates converge to the GNE with the rate of $O(1/t)$ under certain monotonicity assumptions on the problem. The proof follows the techniques in [25], which was established for solving optimization problems from the VI point of view. We start by introducing the monotonicity assumption on the mapping $\mathbf{F}$ over the region $\Omega$.

**Assumption 2.** *The mapping* $\mathbf{F}: \Omega \to \mathbb{R}^{3N}$ *is monotone over* $\Omega$, *i.e.,* $\forall \boldsymbol{\omega}^1, \boldsymbol{\omega}^2 \in \Omega$, *we have*

$$(\boldsymbol{\omega}^1 - \boldsymbol{\omega}^2)^T[\mathbf{F}(\boldsymbol{\omega}^1) - \mathbf{F}(\boldsymbol{\omega}^2)] \geq 0. \quad (15)$$

Hereafter, we will establish the convergence results based on this assumption.

**Remark 2.** [Monotonicity of $\mathbf{F}$] The monotonicity of $\mathbf{F}$ plays an essential role in the proof of standard primal-dual algorithms for solving *optimization* problems [24]. In fact, if the mapping $\mathbf{F}$ is differentiable, then requiring $\mathbf{F}$ to be monotone over the whole primal-dual space $\Omega$ is equivalent to requiring the matrix $\nabla \mathbf{F}(\boldsymbol{\omega}) = \begin{bmatrix} \gamma \mathbf{I} & \mathbf{0} & \tilde{\mathbf{B}}^T \\ \mathbf{0} & \mathbf{H}_h & -\mathbf{I} \\ -\rho\tilde{\mathbf{B}} & \rho\tilde{\mathbf{B}}\mathbf{B}^{-1} & \mathbf{0} \end{bmatrix}$ to be positive semi-definite (PSD), where $\mathbf{H}_h$ is the Hessian of $h$. However, unlike the case in optimization where $\tilde{\mathbf{B}} = \mathbf{B}$, the matrix $\nabla \mathbf{F}(\boldsymbol{\omega})$ here is not *skew-symmetric* as shown in [25] and thus not necessarily PSD even if both functions $f$ and $h$ are convex and $\rho = 1$. This can be mitigated by taking a larger value for $\gamma$, the stronger monotonicity of $\nabla h$, and a proper choice of $\rho$. We will also show from numerical examples that the algorithm can converge to the GNE even when this assumption of monotonicity does not hold.

The GNE $\boldsymbol{\omega}^*$ can be further characterized in the following lemma under Assumption 2, based on Theorem 2.3.5 in [22]. This lemma implies that $\tilde{\boldsymbol{\omega}} \in \Omega$ is an approximate solution of VI$(\Omega, \mathbf{F})$ with accuracy $\epsilon > 0$ if it satisfies

$$(\boldsymbol{\omega} - \tilde{\boldsymbol{\omega}})^T \mathbf{F}(\boldsymbol{\omega}) \geq -\epsilon, \forall \boldsymbol{\omega} \in \Omega. \quad (16)$$

We will show that the sequence $\{\boldsymbol{\omega}^{(t)}\}$ generated by the proposed algorithm satisfies (16).

**Lemma 3.** *Under Assumption 2, the set of GNE $\Omega^*$ can be characterized as*

$$\Omega^* = \bigcap_{\boldsymbol{\omega} \in \Omega} \{\tilde{\boldsymbol{\omega}} \in \Omega : (\boldsymbol{\omega} - \tilde{\boldsymbol{\omega}})^T \mathbf{F}(\boldsymbol{\omega}) \geq 0\}.$$

The updates (13) over the region $\Omega$ can be rearranged in a more compact form

$$\mathbf{v}^{(t+1)} = \arg\min_{\mathbf{v}} f(\mathbf{v}) + \frac{1}{2}\|\mathbf{v} - \mathbf{v}^{(t)}\|_{\beta\mathbf{I}-\rho\tilde{\mathbf{B}}^T\tilde{\mathbf{B}}}^2$$
$$+ \frac{\rho}{2}\|\tilde{\mathbf{B}}\mathbf{v} - \tilde{\mathbf{B}}\mathbf{B}^{-1}\mathbf{q}^{(t)} - \tilde{\mathbf{B}}\mathbf{B}^{-1}\mathbf{w} + \frac{1}{\rho}\boldsymbol{\theta}^{(t)}\|_2^2 \quad (17a)$$

$$\mathbf{q}^{(t+1)} = \arg\min_{\mathbf{q} \in \mathcal{Q}} h(\mathbf{q}) + \frac{\rho}{2}\|\tilde{\mathbf{B}}\mathbf{v}^{(t+1)} - \mathbf{q}$$
$$+ (\mathbf{I} - \tilde{\mathbf{B}}\mathbf{B}^{-1})\mathbf{q}^{(t)} - \tilde{\mathbf{B}}\mathbf{B}^{-1}\mathbf{w} + \frac{1}{\rho}\boldsymbol{\theta}^{(t)}\|_2^2 \quad (17b)$$

$$\boldsymbol{\theta}^{(t+1)} = \boldsymbol{\theta}^{(t)} + \rho\tilde{\mathbf{B}}(\mathbf{v}^{(t+1)} - \mathbf{B}^{-1}\mathbf{q}^{(t+1)} - \mathbf{B}^{-1}\mathbf{w}), \quad (17c)$$

where the measurement $\check{\mathbf{w}}^{(t+1)}$ in (13c) is replaced by the VAR injection $\mathbf{q}^{(t+1)}$ following the power flow equation (3).

For ease of notation, we introduce three matrices

$$\mathbf{H} = \begin{bmatrix} \beta\mathbf{I} - \rho\tilde{\mathbf{B}}^T\tilde{\mathbf{B}} & \mathbf{0} & \mathbf{0} \\ \mathbf{0} & \rho\mathbf{I} & \mathbf{0} \\ \mathbf{0} & \mathbf{0} & \mathbf{I} \end{bmatrix}, \mathbf{M} = \begin{bmatrix} \mathbf{I} & \mathbf{0} & \mathbf{0} \\ \mathbf{0} & \mathbf{I} & \mathbf{0} \\ \mathbf{0} & -\rho\tilde{\mathbf{B}}\mathbf{B}^{-1} & \mathbf{I} \end{bmatrix}$$

$$\mathbf{Q} = \begin{bmatrix} \beta\mathbf{I} - \rho\tilde{\mathbf{B}}^T\tilde{\mathbf{B}} & \mathbf{0} & \mathbf{0} \\ \mathbf{0} & \rho\mathbf{I} & \mathbf{0} \\ \mathbf{0} & -\rho\tilde{\mathbf{B}}\mathbf{B}^{-1} & \mathbf{I} \end{bmatrix}.$$

Note that $\mathbf{Q} = \mathbf{HM}$. We then define a sequence $\{\tilde{\boldsymbol{\omega}}^{(t)}\}$ based on the sequence $\{\boldsymbol{\omega}^{(t)}\}$ where

$$\tilde{\boldsymbol{\omega}}^{(t)} = \begin{bmatrix} \tilde{\mathbf{v}}^{(t)} \\ \tilde{\mathbf{q}}^{(t)} \\ \tilde{\boldsymbol{\theta}}^{(t)} \end{bmatrix} := \begin{bmatrix} \mathbf{v}^{(t+1)} \\ \mathbf{q}^{(t+1)} \\ \boldsymbol{\theta}^{(t)} + \rho\tilde{\mathbf{B}}(\mathbf{v}^{(t+1)} - \mathbf{B}^{-1}\mathbf{q}^{(t)} - \mathbf{B}^{-1}\mathbf{w}) \end{bmatrix}.$$
$$(18)$$

Notice that
$$\boldsymbol{\omega}^{(t+1)} = \boldsymbol{\omega}^{(t)} - \mathbf{M}(\boldsymbol{\omega}^{(t)} - \tilde{\boldsymbol{\omega}}^{(t)}). \quad (19)$$

We then obtain the following lemma that quantifies the discrepancy between $\tilde{\boldsymbol{\omega}}^{(t)}$ and the actual solution to $\mathrm{VI}(\Omega, \mathbf{F})$.

**Lemma 4.** *The sequences $\{\boldsymbol{\omega}^{(t)}\}$ and $\{\tilde{\boldsymbol{\omega}}^{(t)}\}$ satisfy*
$$(\boldsymbol{\omega} - \tilde{\boldsymbol{\omega}}^{(t)})^T \mathbf{F}(\boldsymbol{\omega}) \geq (\boldsymbol{\omega} - \tilde{\boldsymbol{\omega}}^{(t)})^T \mathbf{Q}(\boldsymbol{\omega}^{(t)} - \tilde{\boldsymbol{\omega}}^{(t)}), \forall \boldsymbol{\omega} \in \Omega. \quad (20)$$

**Proof.** We first derive the optimality conditions for the minimization steps (17a) and (17b):
$$(\mathbf{v} - \tilde{\mathbf{v}}^{(t)})^T \left[ \nabla f(\tilde{\mathbf{v}}^{(t)}) + \tilde{\mathbf{B}}^T \tilde{\boldsymbol{\theta}}^{(t)} \right.$$
$$\left. + (\beta \mathbf{I} - \rho \tilde{\mathbf{B}}^T \tilde{\mathbf{B}})(\tilde{\mathbf{v}}^{(t)} - \mathbf{v}^{(t)}) \right] \geq 0, \forall \mathbf{v} \in \mathbb{R}^N \quad (21a)$$
$$(\mathbf{q} - \tilde{\mathbf{q}}^{(t)})^T \left[ \nabla h(\tilde{\mathbf{q}}^{(t)}) - \tilde{\boldsymbol{\theta}}^{(t)} \right.$$
$$\left. + \rho(\tilde{\mathbf{q}}^{(t)} - \mathbf{q}^{(t)}) \right] \geq 0, \forall \mathbf{q} \in \mathcal{Q}. \quad (21b)$$

Also, according to (18), we have that $\forall \boldsymbol{\theta} \in \Theta$,
$$\tilde{\boldsymbol{\theta}}^{(t)} - \boldsymbol{\theta}^{(t)} - \rho \tilde{\mathbf{B}}(\mathbf{v}^{(t+1)} - \mathbf{B}^{-1}\mathbf{q}^{(t)} - \mathbf{B}^{-1}\mathbf{w}) = \mathbf{0}. \quad (22)$$

Combining (21) and (22), we have that $\forall \boldsymbol{\omega} \in \Omega$,
$$(\boldsymbol{\omega} - \tilde{\boldsymbol{\omega}}^{(t)})^T \mathbf{F}(\tilde{\boldsymbol{\omega}}^{(t)}) \geq (\boldsymbol{\omega} - \tilde{\boldsymbol{\omega}}^{(t)})^T \mathbf{Q}(\boldsymbol{\omega}^{(t)} - \tilde{\boldsymbol{\omega}}^{(t)}). \quad (23)$$

Recalling that $\mathbf{Q} = \mathbf{HM}$ and $\mathbf{F}(\boldsymbol{\omega})$ is monotone over $\Omega$, we further obtain (20). ∎

The right hand side of (20) can be handled using the following lemma.

**Lemma 5.** *The sequences $\{\boldsymbol{\omega}^{(t)}\}$ and $\{\tilde{\boldsymbol{\omega}}^{(t)}\}$ satisfy $\forall \boldsymbol{\omega} \in \Omega$,*
$$(\boldsymbol{\omega} - \tilde{\boldsymbol{\omega}}^{(t)})^T \mathbf{HM}(\boldsymbol{\omega}^{(t)} - \tilde{\boldsymbol{\omega}}^{(t)}) \quad (24)$$
$$= \frac{1}{2}(\|\boldsymbol{\omega} - \boldsymbol{\omega}^{(t+1)}\|_{\mathbf{H}}^2 - \|\boldsymbol{\omega} - \boldsymbol{\omega}^{(t)}\|_{\mathbf{H}}^2) + \frac{1}{2}\|\boldsymbol{\omega}^{(t)} - \tilde{\boldsymbol{\omega}}^{(t)}\|_{\mathbf{R}}^2,$$
*where $\mathbf{R} = 2\mathbf{HM} - \mathbf{M}^T \mathbf{HM}$.*

**Proof.** By the relation (19), we have
$$\mathbf{M}(\boldsymbol{\omega}^{(t)} - \tilde{\boldsymbol{\omega}}^{(t)}) = \boldsymbol{\omega}^{(t)} - \boldsymbol{\omega}^{(t+1)}. \quad (25)$$

Applying the fact that
$$(\mathbf{a} - \mathbf{b})^T \mathbf{H}(\mathbf{c} - \mathbf{d}) = \frac{1}{2}(\|\mathbf{a} - \mathbf{d}\|_{\mathbf{H}}^2 - \|\mathbf{a} - \mathbf{c}\|_{\mathbf{H}}^2)$$
$$+ \frac{1}{2}(\|\mathbf{c} - \mathbf{b}\|_{\mathbf{H}}^2 - \|\mathbf{d} - \mathbf{b}\|_{\mathbf{H}}^2),$$

we have
$$(\boldsymbol{\omega} - \tilde{\boldsymbol{\omega}}^{(t)})^T \mathbf{H}(\boldsymbol{\omega}^{(t)} - \boldsymbol{\omega}^{(t+1)})$$
$$= \frac{1}{2}(\|\boldsymbol{\omega} - \boldsymbol{\omega}^{(t+1)}\|_{\mathbf{H}}^2 - \|\boldsymbol{\omega} - \boldsymbol{\omega}^{(t)}\|_{\mathbf{H}}^2)$$
$$+ \frac{1}{2}(\|\boldsymbol{\omega}^{(t)} - \tilde{\boldsymbol{\omega}}^{(t)}\|_{\mathbf{H}}^2 - \|\boldsymbol{\omega}^{(t+1)} - \tilde{\boldsymbol{\omega}}^{(t)}\|_{\mathbf{H}}^2). \quad (26)$$

Additionally, we can also obtain from (19) that
$$\|\boldsymbol{\omega}^{(t)} - \tilde{\boldsymbol{\omega}}^{(t)}\|_{\mathbf{H}}^2 - \|\boldsymbol{\omega}^{(t+1)} - \tilde{\boldsymbol{\omega}}^{(t)}\|_{\mathbf{H}}^2 \quad (27)$$
$$= \|\boldsymbol{\omega}^{(t)} - \tilde{\boldsymbol{\omega}}^{(t)}\|_{\mathbf{H}}^2 - \|\boldsymbol{\omega}^{(t)} - \tilde{\boldsymbol{\omega}}^{(t)} - \mathbf{M}(\boldsymbol{\omega}^{(t)} - \tilde{\boldsymbol{\omega}}^{(t)})\|_{\mathbf{H}}^2$$
$$= \|\boldsymbol{\omega}^{(t)} - \tilde{\boldsymbol{\omega}}^{(t)}\|_{\mathbf{R}}^2,$$

where
$$\mathbf{R} := 2\mathbf{HM} - \mathbf{M}^T \mathbf{HM} = 2\mathbf{Q} - \mathbf{M}^T \mathbf{Q}$$
$$= \begin{bmatrix} \beta \mathbf{I} - \rho \tilde{\mathbf{B}}^T \tilde{\mathbf{B}} & 0 & 0 \\ 0 & \rho(\mathbf{I} - \rho \mathbf{B}^{-T} \tilde{\mathbf{B}}^T \tilde{\mathbf{B}} \mathbf{B}^{-1}) & \rho \mathbf{B}^{-T} \tilde{\mathbf{B}}^T \\ 0 & -\rho \tilde{\mathbf{B}} \mathbf{B}^{-1} & \mathbf{I} \end{bmatrix}.$$

By combining (26) and (27), we obtain (24), which concludes the proof. ∎

Note that $\mathbf{R}$ is skew-symmetric. With large enough $\beta > 0$ and small enough $\rho > 0$, both submatrices $\beta \mathbf{I} - \rho \tilde{\mathbf{B}}^T$ and $\rho(\mathbf{I} - \rho \mathbf{B}^{-T} \tilde{\mathbf{B}}^T \tilde{\mathbf{B}} \mathbf{B}^{-1})$ can be PSD, leading $\mathbf{R}$ to be PSD as well. With this observation, we are ready to present the main result about the convergence of the algorithm under Assumption 2.

**Theorem 2.** *Under Assumption 2, if the parameters $\rho > 0$ and $\beta > 0$ satisfy: i) $\rho \leq 1/\|\mathbf{B}^{-T} \tilde{\mathbf{B}}^T \tilde{\mathbf{B}} \mathbf{B}^{-1}\|_2$; ii) $\beta \geq \rho \cdot \|\tilde{\mathbf{B}}^T \tilde{\mathbf{B}}\|_2$, then given the sequences $\{\boldsymbol{\omega}^{(\tau)}\}$ and $\{\tilde{\boldsymbol{\omega}}^{(\tau)}\}$ and letting*
$$\underline{\tilde{\boldsymbol{\omega}}}^{(t)} = \frac{1}{t+1} \sum_{\tau=0}^{t} \tilde{\boldsymbol{\omega}}^{(\tau)}, \quad (28)$$

*we have, $\forall t > 0$, $\underline{\tilde{\boldsymbol{\omega}}}^{(t)} \in \Omega$ and*
$$(\underline{\tilde{\boldsymbol{\omega}}}^{(t)} - \boldsymbol{\omega})^T \mathbf{F}(\boldsymbol{\omega}) \leq \frac{1}{2(t+1)} \|\boldsymbol{\omega} - \boldsymbol{\omega}^{(0)}\|_{\mathbf{H}}^2, \forall \boldsymbol{\omega} \in \Omega. \quad (29)$$

**Proof.** First, due to the convexity of $\Omega$, (28) implies that $\underline{\tilde{\boldsymbol{\omega}}}^{(t)} \in \Omega$. Second, with the conditions in the Theorem, the parameters $\rho$ and $\beta$ make the matrix $\mathbf{H}$ and $\mathbf{R}$ both PSD. Therefore by combining (20) and (24), we obtain that $\forall \tau > 0$,
$$(\boldsymbol{\omega} - \tilde{\boldsymbol{\omega}}^{(\tau)})^T \mathbf{F}(\boldsymbol{\omega}) + \frac{1}{2}(\|\boldsymbol{\omega} - \boldsymbol{\omega}^{(\tau)}\|_{\mathbf{H}}^2 - \|\boldsymbol{\omega} - \boldsymbol{\omega}^{(\tau+1)}\|_{\mathbf{H}}^2)$$
$$\geq \|\boldsymbol{\omega}^{(t)} - \tilde{\boldsymbol{\omega}}^{(t)}\|_{\mathbf{R}}^2 \geq 0, \forall \boldsymbol{\omega} \in \Omega,$$

i.e., $\forall \boldsymbol{\omega} \in \Omega$,
$$(\boldsymbol{\omega} - \tilde{\boldsymbol{\omega}}^{(\tau)})^T \mathbf{F}(\boldsymbol{\omega}) + \frac{1}{2}\|\boldsymbol{\omega} - \boldsymbol{\omega}^{(\tau)}\|_{\mathbf{H}}^2 \geq \frac{1}{2}\|\boldsymbol{\omega} - \boldsymbol{\omega}^{(\tau+1)}\|_{\mathbf{H}}^2.$$

By summing up the inequality over $\tau \geq 0$, we obtain
$$\left((t+1)\boldsymbol{\omega} - \sum_{\tau=0}^{t} \tilde{\boldsymbol{\omega}}^{(\tau)}\right)^T \mathbf{F}(\boldsymbol{\omega}) + \frac{1}{2}\|\boldsymbol{\omega} - \boldsymbol{\omega}^{(0)}\|_{\mathbf{H}}^2 \geq 0,$$

and equivalently we have
$$(\underline{\tilde{\boldsymbol{\omega}}}^{(t)} - \boldsymbol{\omega})^T \mathbf{F}(\boldsymbol{\omega}) \leq \frac{1}{2(t+1)}\|\boldsymbol{\omega} - \boldsymbol{\omega}^{(0)}\|_{\mathbf{H}}^2, \forall \boldsymbol{\omega} \in \Omega, \quad (30)$$

which concludes the proof. ∎

Theorem 2 shows that after $t$ iterations of the updates (17), the point $\underline{\tilde{\boldsymbol{\omega}}}^{(t)}$ is an approximate solution to $\mathrm{VI}(\Omega, \mathbf{F})$ with accuracy $O(1/t)$ (recall (16)), which establishes the convergence of the algorithm.

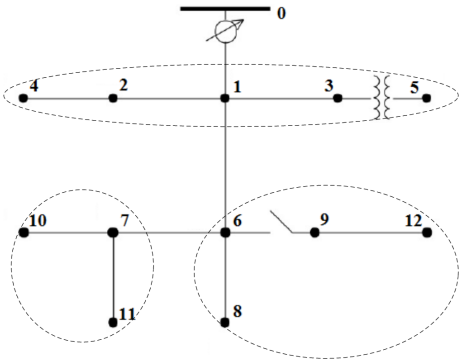

Fig. 2. The IEEE 13-bus feeder case for distribution systems. The dashed circles represent the partition of buses as areas under only a locally connected communication network.

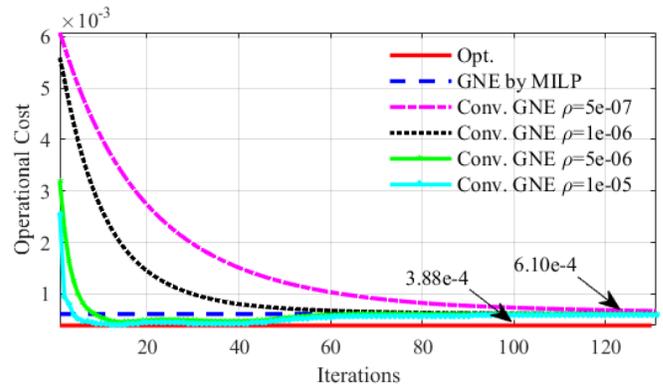

Fig. 3. The operational cost of the global optimum, the GNE computed by solving the MILP, and the GNE attained by the proposed algorithm with various values of $\rho$.

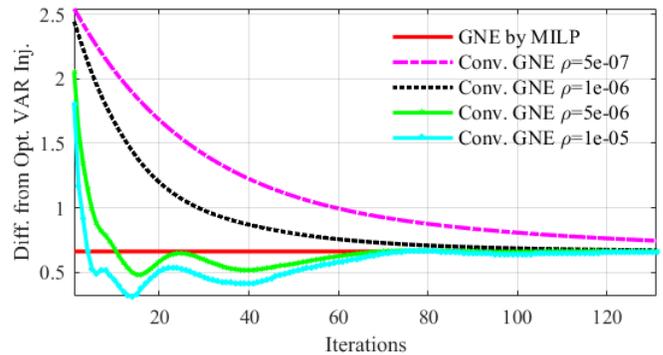

Fig. 4. The difference ($\ell_2$-norm) of the optimal VAR injection and the injections at the GNE attained by the proposed algorithm using various values of $\rho$.

## V. NUMERICAL SIMULATIONS

In this section, we evaluate the performance of voltage control under local communications via numerical simulations. The IEEE 13-bus feeder case for distribution systems [26] are tested, where the partition of buses into communication areas is illustrated in Fig. 2. The line impedances are assumed to be uniform with $0.233 + j0.366\Omega$ under the base of 4.16 kV and 100 kVA. We choose $\gamma = 1$, the desired voltage $\boldsymbol{\mu} = \mathbf{1}$, and quadratic cost function $C_j(q_j) = c_j/2 \|q_j\|_2^2$ with identical $c_j = 10^{-4}, \forall j \in \mathcal{N}_p$. In this case, the mapping $\boldsymbol{\Phi}$ defined in (9) becomes a linear mapping and is strongly monotone with the smallest eigenvalue $9.85 \times 10^{-5}$. Hence by Lemma 2 the GNE is unique in this example. The feasible set of VAR injection at each bus is $\mathcal{Q}_j = [-0.8, 0.8]$ p.u..

We first compare the performance of the GNE and the global optimum of the problem $\mathcal{P}_0$ when the communication network is strongly-connected. To this end, we solve the equilibrium conditions for the GNE (8) directly, by transforming the complementary slackness equations into a mixed integer linear programming (MILP) as the technique developed in [27]. The MILP can be attacked readily by standard solvers as Gurobi [28]. As shown in Fig. 3, in contrast to the global optimization result, the equilibrium has global objective values that are 1.57 times greater. This loss of efficiency characterizes the value of communication links in voltage control.

We then implement the proposed algorithm at all buses with only neighbor-to-neighbor information exchange. The value of $\beta$ is chosen to be $\rho \cdot \|\tilde{\mathbf{B}}^T \tilde{\mathbf{B}}\|_2$. As illustrated in Fig. 3, the equilibrium-learning algorithm successfully converges to the unique GNE as solved by the MILP approach with sublinear rate as proved. The greater the value $\rho$ is, the faster the algorithm converges. Nonetheless, $\rho$ has to be small enough to avoid oscillation or possible divergence. Fig. 4 shows the difference ($\ell_2$-norm) between the convergent VAR injection at the GNE and the optimal one. The same convergence as for objective values is exhibited for VAR injections. Note that in this example, the algorithm delivers desired convergence even without Assumption 2 since the matrix $\nabla \mathbf{F}$ has negative smallest eigenvalue.

We also compare the convergence speed of the proposed algorithm and the extra-gradient algorithm whose global convergence requires the weakest monotonicity assumption on the mapping $\mathbf{F}$, see Remark 1. We vary the value of $c_j$ to vary the monotonicity of $\mathbf{F}$. The numbers of iterations needed for convergence to $\|\mathbf{q}^{(t)} - \mathbf{q}^*\| \le \epsilon = 10^{-8}$ are listed in Table I. The best parameters such as the step-size for the EG algorithm or $\rho$ and $\beta$ for our algorithm, are chosen and fixed for all test cases. '$-$' represents the case the algorithm either ges or takes more than 500000 iterations to converge. It is shown that the EG algorithm is much slower, or even fails to converge, compared to our algorithm, which numerically corroborates the applicability of our algorithm under unfavorable monotonicity conditions of $\mathbf{F}$.

TABLE I
THE NUMBERS OF ITERATIONS NEEDED FOR THE EG AND OUR ALGORITHM TO CONVERGE UNDER VARIOUS MONOTONICITY CONDITIONS OF $\mathbf{F}$ DETERMINED BY $c_j$.

| $c_j$ | $10^{-4}$ | $10^{-2}$ | $10^{-1}$ | 1 |
|---|---|---|---|---|
| EG Alg. | - | 456841 | 88536 | 5089 |
| Our Alg. | 844 | 620 | 524 | 422 |

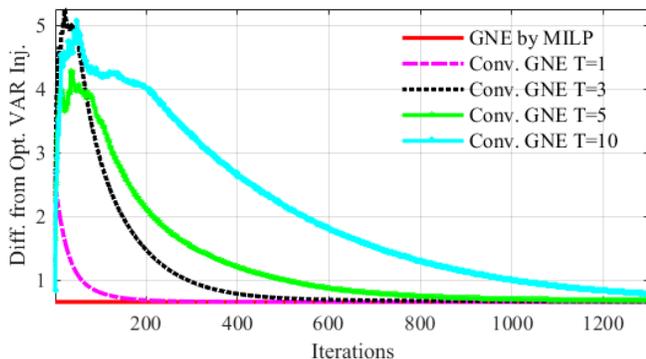

Fig. 5. The difference ($\ell_2$-norm) of the optimal VAR injection and the injections at the GNE attained by the proposed algorithm with various update delay bounds $T$.

In addition, we have also numerically investigated the performance of asynchronous updates with bounded update delays caused by intermittent communication within each area. As the model in [11], we assume each area conducts the update at least every $T$ iterations. As shown in Fig. 5, the proposed GNE learning algorithm is robust to this asynchronous update.

## VI. CONCLUSIONS AND FUTURE WORK

The performance of voltage control using DERs is limited by the under-deployed communication infrastructure in most of the existing power distribution systems. In the present work, we propose a game-theoretic characterization of voltage control with only locally connected communication networks. DERs are partitioned into communication areas and information exchange is allowed only within each area. Existence and uniqueness conditions for the equilibrium are analyzed, followed by the design of a fully distributed equilibrium-learning algorithm. We have shown that under a certain monotonicity assumption, the algorithm converges to the GNE at the rate of $O(1/t)$. Numerical tests are also presented to verify its superior convergence property.

As future work, it would be interesting to analyze how the structure of the communication network impacts the efficiency of the GNE compared with the global optimum, so as to further understand the fundamental value of communication links in voltage control. It is also worth attempting to establish convergence results under weaker assumptions on the monotonicity of $\mathbf{F}$.